\documentclass[aps,prb,groupedaddress]{revtex4}
\usepackage{graphicx}
\begin{document}
\title{Continuous and correlated nucleation during nonstandard island growth at Ag/Si(111)-7$\times$7 heteroepitaxy}
\author{P. Koc\'an}
\author{P. Sobot\'{\i}k}
\email[]{Pavel.Sobotik@mff.cuni.cz}
\author{I. O\v{s}t'\'adal}

\affiliation{Department of Electronics and Vacuum Physics, Faculty of Mathematics and Physics,
Charles University, V Hole\v{s}ovi\v{c}k\'ach 2, 180 00 Praha 8, Czech Republic}
\author{M. Kotrla}
\affiliation{Institute of Physics, Academy of Sciences of the Czech Republic, Na Slovance 2, 182 21 Praha 8, Czech Republic}
\date{\today}

\begin{abstract}
We present a combined experimental and theoretical study of submonolayer
heteroepitaxial growth of Ag on Si(111)-$7 \times 7$ at temperatures
from 420 K to 550 K when Ag atoms can easily diffuse on the surface and
the reconstruction $7 \times 7$  remains stable.
STM measurements for coverages from 0.05 ML to 0.6 ML show that there
is an excess of smallest islands (each of them fills up just one
half-unit cell -- HUC) in all stages of growth. Formation of 2D wetting
layer proceeds by continuous nucleation of the smallest islands in the proximity of
larger 2D islands (extended over several HUCs) and following coalescence with them. Such a growth scenario is verified by
kinetic Monte Carlo simulation which uses a coarse-grained model
based on a limited
capacity of HUC and a mechanism which increases nucleation probability in a neighbourhood of already saturated HUCs (correlated nucleation). The model provides a good fit for experimental
dependences of the relative number of Ag-occupied HUCs and the preference in
occupation of
faulted HUCs on temperature and amount of deposited Ag.
Parameters obtained for the hopping of Ag adatoms between HUCs agree
with those reported earlier for initial stages of growth.
The model provides two new parameters - maximum number of Ag atoms inside
HUC, and on HUC boundary.
\end{abstract}

\maketitle

\section{Introduction}
Heteroepitaxial growth of metals on silicon surfaces has been studied for decades
and morphologies of various grown structures were reported \cite{Tanishiro,
Custance, Nedder, Vitali, Susc2000, nase}. Reconstruction of oriented
semiconductor surfaces determines the mobility of deposited adatoms and substantially
influences growth mechanism. Recently reported experimental studies on
self-organised growth of arrays of ordered metal islands -- quantum dots -- on the
Si(111)-7$\times$7 surface \cite{Vitali, Li} stimulate need of
detailed
understanding of mechanisms controlling the growth. Heteroepitaxy of Ag on the
Si(111)-7$\times$7 surface represents one of frequently studied problems due to
nonreactivity of Ag with the reconstructed surface, abrupt interface and
negligible interdiffusion of both elements. The growth mode is of the
Stranski-Krastanov type -- 3D islands are formed on a 2D Ag transition layer
(wetting layer) grown on the 7$\times$7 silicon surface.

Our initial STM study
 of Ag/Si heteroepitaxy at low coverage
\cite{Susc2000} showed a growth
mechanism affected by trapping Ag adatoms in triangular units of the 7$\times$7
reconstruction -- half unit cells (HUCs). The HUCs are of two types: 'faulted',
FHUC (containing a structural fault according to the DAS model \cite{Takayanagi})
and 'unfaulted', UHUC. For deposited Ag atoms, the two types of HUCs represent 
potential wells with different depths $E_F>E_U$. This
leads to preferential nucleation in the FHUCs (the preference $P_F$ is defined as
a ratio of FHUCs containing Ag adatoms to all occupied HUCs). In the successive
work \cite{Pepa} we investigated processes of adatom diffusion, nucleation and
island formation at the beginning of the Ag growth for a deposited amount $<0.1$~ML
(1 ML $\approx 7.83 \times 10^{14}$ atoms/cm$^2$), both experimentally and
theoretically. We developed and used a simple coarse-grained model for the kinetic
Monte Carlo (KMC) simulations. Fitting of experimental data provided
values $E_F \approx E_U = (0.75 \pm 0.10)$ eV, $E_F -E_U < 0.05$ eV and frequency
prefactors $\nu^{0}_{F} \approx \nu^{0}_{U} = 5 \times 10^{(9 \pm 1)} s^{-1}$.

Recently, we extended our STM measurements to coverages up
to 0.6~ML\cite{nase}.
In this regime
a discontinuous 2D film -- wetting layer \cite{2D3D} -- is formed.
We observed large 2D islands completely covering several HUCs. We did  not observe any island to overgrow the HUC boundaries at its perimeter. 
This results in triangularly jagged island shapes. There was a considerable number of stable
Ag clusters - each of them formed {\it inside} a HUC.
We denoted such islands as 1-HUC islands.
These islands were observed for coverages up to 0.6~ML
and for high temperatures (540 K) as well.
Statistical analysis of island population on the
surface revealed a number of 1-HUC islands much higher than
a value expected by the "standard" model of island film growth\cite{Venables} 
in which island density saturates and then all 
adatoms are captured by existing islands.

We suggested a possible growth mechanism compatible with our observations: A
single 1-HUC island grows by adatom capturing until  a maximum number of adatoms
which can be accommodated in a HUC is reached - an 1-HUC island is saturated. The saturated and isolated island does not capture diffusing adatoms any more. It leads to an increase of Ag adatom concentration around such islands and results in
enhanced nucleation of new islands in proximity of saturated HUCs (correlated nucleation). Larger islands
grow by coalescence of smaller saturated islands.

In this paper, we verify the above nonstandard growth scenario by
 a combined experimental and theoretical study.
We performed STM experiments in a temperature
range when Ag atoms can easily diffuse on the surface and
the structure $7 \times 7$  remains stable.
We measured dependences of several structure related quantities (object densities, preferences of occupation, island size distribution)
on deposited amount, substrate temperature and deposition flux.
We show that the results of measurements can
be explained by KMC simulation using a
modified
coarse-grained model
developed from the one we used for low coverage growth \cite{Pepa}.
The new model, which takes into account limited capacity of HUCs and
correlated nucleation, explains both morphological and statistical
properties of island
growth and coalescence on the reconstructed Si(111)-$7 \times 7$ surface.

\section{Experiments}
Series of samples with various amounts of deposited Ag from 0.05~ML to 0.6~ML
were prepared at temperatures from 420~K to 550~K at deposition rate
$F_1= 0.011$ MLs$^{-1}$. Another series with Ag amounts from 0.05~ML to
0.3~ML were deposited at $T = (492 \pm 10)$~K and
deposition rate $F_2 = 0.0005$~MLs$^{-1}$.
Ag was evaporated from a tungsten filament in an ultra-high vacuum chamber,
the deposited amount was measured by a quartz thickness monitor with
an absolute
accuracy of $\pm$~10~\%. Sb doped Si(111) substrates with a miscut of $\pm$~0.1$^{\circ}$
and resistivity of 0.005 -- 0.01~$\Omega$cm were heated by passing dc current
(temperature calibrated with accuracy of $\pm$~10~K). Other experimental details
(substrate treatment, thickness monitor and temperature calibration etc.) have
been already reported elsewhere \cite{PhD,Pepa}.
Before STM measurements, deposited films relaxed at least 1 hour at room temperature (RT). Experimental procedures were performed at pressure
$< 2 \times 10 ^{-8}$~Pa. We used an STM of our design and construction with
electrochemically polished tungsten tips.

\section{Experimental results}
\label{ExpRes}

\begin{figure}
\includegraphics[width=7cm]{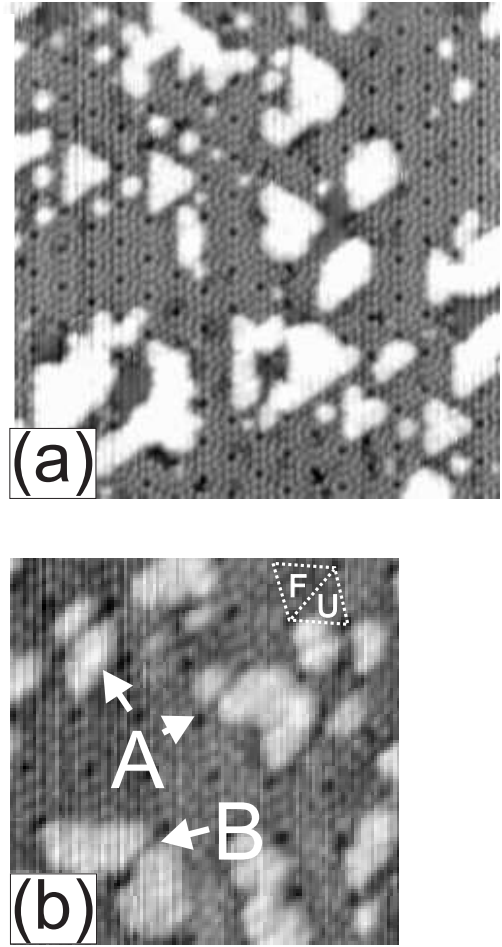}
\caption{(a) STM image of island morphology -- white objects are Ag
islands, with the exception of the smallest dots (impurities),
$35\times35$ nm$^2$ area, evaporated amount
$d=(0.5\pm0.1$)~ML at temperature $T=490$~K and flux
$F=0.0005$~MLs$^{-1}$; (b) detail of island morphology, $16\times16$ nm$^2$ area: A - the 1-HUC island grown in a HUC adjacent to a larger island, B - two
larger separated islands.} \label{notcon}
\end{figure}

\begin{figure}
\includegraphics[width=8cm]{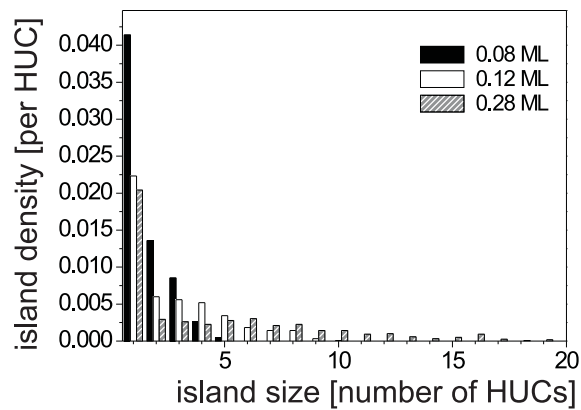}
\caption{
Island size distribution for three different coverages. Samples were prepared at temperature 490~K and at deposition
rate 0.011~MLs$^{-1}$.
}
\label{distr}
\end{figure}

Fig. \ref{notcon}a shows an example of  morphology  of Ag islands grown at substrate temperature 490 K as observed in STM.
Following morphological features has
been found by analysis of a large number of images taken from various samples:
\begin{itemize}
\item [(i)] Ag forms 2D islands of various sizes bordered always by dimer rows of the $7
\times 7$ reconstruction. The dimer rows at island boundary are not filled by
Ag atoms, dimer rows inside larger islands are filled (overgrown) by Ag atoms --
see detail A on Fig. \ref{notcon}b.
\item [(ii)]  We often observed islands covering
adjacent HUCs but clearly separated by the dimer row - see detail B on Fig.
\ref{notcon}b. 
\item [(iii)] Triangular shapes
of larger islands follow the orientation of FHUCs (see Fig. \ref{notcon}a and
Ref. \onlinecite{nase}).
\item [(iv)] An important feature is excess
of 1-HUC islands in island size distribution even at very low deposition rates, higher temperatures and coverages up to 0.6 ML \cite{magic}. 
This is illustrated in  Fig. \ref{distr} which shows the island size 
distribution for different coverages
(see also Fig. 1 and Fig. 2 in Ref. \onlinecite{nase}).
The basic size unit is an area of one HUC.
The 1-HUC islands clearly dominate in all distributions.

The 1-HUC
islands grow preferentialy in FHUCs and in proximity of larger islands rather than in vacant areas
of the surface (Fig. \ref{notcon}a). The formation of the wetting layer proceeds as continuous
nucleation of new 1-HUC islands.
\item [(v)] STM imaging at room temperature does not allow to distinguish number of Ag atoms contained in
the 2D Ag island \cite {dif} with exception of the smallest objects -- HUCs containing 1 or 2 atoms.
\end{itemize}
Following quantities were obtained by statistical analysis of
STM images: preference $P_F$; island size distribution (island size is measured in 
numbers of HUCs covered by the island); total coverage $\theta$ --
the relative number of occupied HUCs (ratio of the occupied HUCs
to the all HUCs on the surface),
and 1-HUC coverage $\theta_{\rm 1-HUC}$  -- 
the relative number of 1-HUC islands (defined similarly).
Ag objects are nonempty HUCs as well as islands larger than a HUC.
When no Ag islands larger than a
HUC are present $\theta = n_{Ag} $, where $n_{Ag}$ is density of Ag objects ($n_{Ag}$ number of objects normalized to the number of all HUCs on the surface.
The preference $P_F$ reflects existence of two different
potential wells $E_F > E_U$ on the surface.

\begin{figure}
\includegraphics[width=8cm]{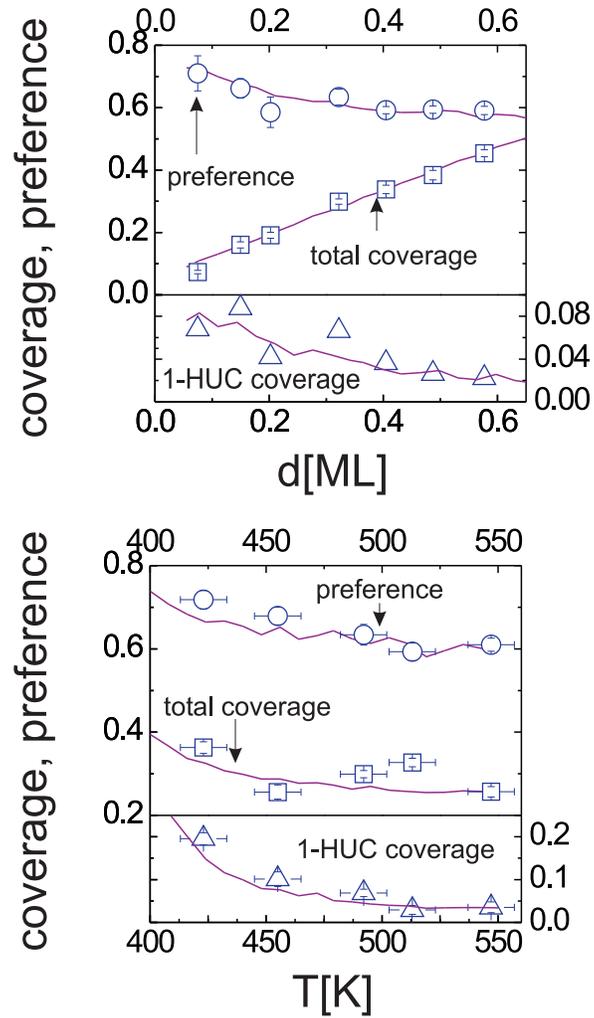}
\caption{Dependence of total coverage, 1-HUC coverage,  and preference on amount of
deposited Ag at $T=490$~K (upper panel) and on substrate temperature for
$d=0.3$~ML (lower panel). Symbols - experimental data, lines - best fit using the
model with centere and boundary areas. All samples were prepared at deposition rate
$F_1=0.011$~MLs$^{-1}$. }
\label{fastdep}
\end{figure}

We measured dependences of $P_F,  \theta$ and $\theta_{\rm 1-HUC}$ 
on deposited amount $d$ 
(upper panel in  Fig. \ref{fastdep})
and substrate temperature 
 (lower panel in  Fig. \ref{fastdep}).
In the studied
range of deposition parameters the 
total coverage $\theta$
is proportional to the deposited amount of Ag and 
decreases only slightly with the increasing
substrate temperature. When islands larger than 1-HUC begin to grow, preference
$P_F$ decreases
with both the deposited amount and the substrate temperature.
However, it remains
larger than 0.6 due to continuous nucleation of new 1-HUC islands
preferably in FHUCs.

\section{Simulation models}
\subsection{Original model}

A coarse-grained KMC model with an algorithm derived from Ref.
\onlinecite{Clarke} was successfully applied for simulation of
early stages of nucleation \cite{Pepa}. The model uses HUCs as
basic units of the surface. Events included in the model
correspond to the following growth scenario: Ag atoms arrive at
the surface in random positions with a rate given by flux F.
Diffusion of Ag adatoms on the Si substrate is modelled by
thermally activated hops to neighboring HUCs. Depending on the HUC type, there are two different contributions to activation energy
from interaction with the substrate.
A frequency prefactor $\nu_0$ is assumed (for simplicity)
to be the same for FHUCs and UHUCs. 
The transient mobility of impinging Ag adatoms was included into the model to explain and simulate  the short-range ordering of Ag objects and the low value of the total coverage at temperatures too low for sufficient adatom mobility between HUCs \cite{Vasco1,Vasco2} (however,  in a temperature range, when Ag atoms can easily diffuse on the surface, this mechanism is not much important for the grown morphologies).
 Hopping adatoms can create nuclei inside HUCs. Let $n$ be a number
of Ag atoms in a HUC. The model assumes existence of a critical
nucleus size $n^{*}$. Nuclei of more than $n^{*}$ Ag atoms are
stable. Nuclei with the size $n\leq n^{*}$  can
decay with activation energy proportional to $n$. Hopping rate of
an Ag atom out of a HUC is approximated as $\nu_n^{F, U} = n
\nu_0 \exp \{ - [E_{F, U}+(n-1)E_a ]/k_B T\}$, where $E_a$ represents
effective Ag-Ag interaction and $k_B$ is Boltzmann's constant.  The values $ E_F \approx E_U = (0.75
\pm 0.10)$ eV, $E_F -E_U < 0.05$ eV, $E_a \approx 0.05$ eV,
critical nucleus size $n^{*} = 5$, and the frequency prefactor for
hopping out of HUCs, $\nu_{0} \ = 5 \times 10^{(9 \pm 1)} s^{-1}$,
were obtained in the Ref. \onlinecite{Pepa}.
The original model has been used only for low coverages where real
limits of capacity  of HUCs cannot be reached.  HUCs were
treated like potential wells with unlimited capacity. When higher
amount of Ag is deposited this simplification has to be replaced
by a certain restriction.

\subsection{Model with a simple constraint}
The simplest way how to introduce a limitation of capacity  is its
direct implementation within a "standard" growth scenario
\cite{Venables} for growth simulations. At the beginning of growth
the density of Ag nuclei reaches a saturated value and the islands
grow simply by capturing hopping adatoms. Each HUC can
accommodate $n_H$ Ag atoms at the most and the next deposited or
diffusing atom is forced to sit to the nearest HUC occupied by less than
$n_H$ Ag atoms. Hence, an island containing more than $n_H$
adatoms overgrows HUC boundaries
and extends over neighboring HUCs. 

\begin{figure}
\includegraphics[width=8cm]{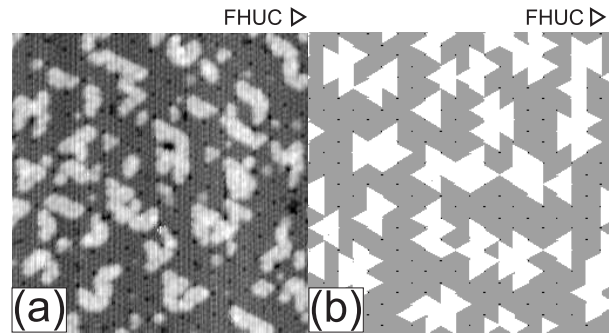}
\caption{(a) STM image of $35\times35$ nm$^2$ area, deposited amount
$d=(0.50\pm0.05)$~ML at temperature $T=490$~K and flux $F=0.011$~MLs$^{-1}$; (b)
layer simulated under the same conditions using "standard" 2D growth scenario (model with a simple constraint). \\
 } \label{clasexp}
\end{figure}

\begin{figure}
\includegraphics[width=8cm]{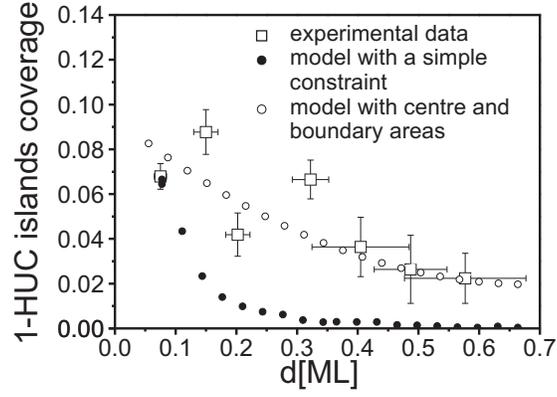}
\caption{
Comparison of the dependence of  1-HUC island coverage
on deposited amount  at temperature
$T=490$~K and flux $F=0.011$~MLs$^{-1}$ in experiment and in two different models.
}
\label{singlcovr}
\end{figure}

We tried to fit experimental data using the above simple modification.
The same model and parameter values as in Ref. \onlinecite{Pepa} were used, only limited capacity of a HUC, $n_H$, was included.
The value $n_H$ was determined by fitting the experimentally
obtained dependence of the relative number of occupied HUCs,
$\theta$, on deposited amount $d$. The fit provided the value of $n_H
= 45 \pm 5$. However, the morphology obtained by using "standard" mechanism differs from the experimental data -- Fig. \ref{clasexp}. Only a few 1-HUC islands are visible in the simulated
layer. A statistical analysis of the experimental results reveals
excess of 1-HUC islands in comparison with the simulated growth --
Fig. \ref{singlcovr}.
Therefore, the concept  of limited capacity needs to be
implemented in a more subtle way taking into account the role of
1-HUC islands in agreement with the scenario suggested in Ref.
\onlinecite{nase} and detailed experimental observations presented in Sec. \ref{ExpRes}.

\subsection{Model with centre and boundary areas - final model}
During the growth a 1-HUC island captures more and more diffusing
adatoms until its size reaches a saturated value $n_S$ given by
the limited capacity of the HUC. The saturated 1-HUC Ag island
cannot further grow by adatom capture. The value $n_S$ is assumed
to be the same for islands in both types of HUCs. When a
diffusing adatom meets the saturated HUC, it hops fast out leaving dimer
row unoccupied. This
event is simulated by means of setting the energy barrier for
hopping out of saturated HUCs close to zero value.

Larger islands grow by coalescence with 1-HUC islands only by the
following mechanism: Boundaries between adjacent HUCs can be
filled only if the cells are saturated. Such two HUCs can coalesce
with assistance of a certain number of hopping adatoms, $n_B$,
completing each HUC. In the simulation, the adjacent saturated
HUCs can incorporate new adatoms until the boundary areas in HUCs
(dimer rows separating adjacent HUCs) are filled by Ag atoms. The
maximum number of atoms in a HUC with all the boundary areas
filled is then $n_S + 3 \times n_B$. Therefore, in the computer
model, each HUC is formally divided into one centre and three
boundary areas. The boundary area can accept Ag atoms if the
centre areas of both adjacent HUCs are saturated -- each contains
$n_S$ Ag atoms (i.e. a saturated 1-HUC island).

Each hopping event in the computer model is selected with
probability determined by activation energy calculated with
respect to a number of Ag atoms at a given position (HUC). We
started with  simulations in which the hopping probability was not
affected by occupancy of a destination site (HUC). These
simulations provided high population of 1-HUC islands but fitting
of model parameters failed in achieving quantitative agreement
with
the experimentally measured dependences.
The concentration of 1-HUC islands obtained by the simulations
was much higher than the experimentally observed value.
The experimentally observed increase of
1-HUC island density
in proximity
of larger islands was not reproduced by the model.

The enhanced nucleation of a new island in proximity of the
saturated one (in an adjacent HUC) can be explained physically by
an increase of time spent by a diffusing Ag adatom in the close
neighborhood of saturated HUC. This might be caused by an
interaction of Ag adatom with Ag atoms in saturated HUC. Another reason may
be a change of the barrier for diffusion over the boundary of
saturated HUC due to relaxation of Si atoms. The effect is
modeled by decrease of the barrier for hopping into the saturated
HUC by $\Delta E_S$.
It increases probability of nucleation near the saturated island.
The correlated nucleation would also imply decrease of the
concentration of 1-HUC islands because during further growth more
1-HUC islands will join larger islands.
This modification implies the need to use a
model with hopping rates depending on a final position. The
introduction of anisotropy for hopping makes the code
technically a bit more complicated
(a hop-oriented code has to be employed).

\section{Simulation results and discussion}

\begin{figure}
 \includegraphics[width=8cm]{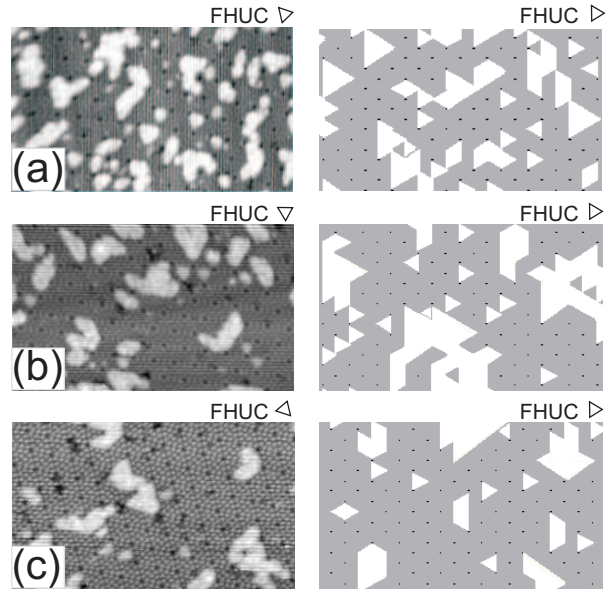}
 \caption{Examples of experimental (left column) and simulated
 (right column) morphology of films
with $d=(0.32 \pm 0.03)$~ML obtained by
deposition at two
temperatures and two deposition rates  (a) $T=490$~K, $F_1=0.011~$MLs$^{-1}$;
(b) $T=550$~K.
$F_1=0.011~$MLs$^{-1}$; (c)   $T=490$~K,
 $F_2=0.0005~$MLs$^{-1}$. Area $35\times20$ nm$^2$. The model with centre and boundary areas was used.
\label{fdmorph}}
\end{figure}

Simulations with the final model (the model with centre and boundary areas) reproduce well growth
morphologies observed in the experiment.
 Figures
\ref{fdmorph}a and \ref{fdmorph}b show examples of
real and simulated growth morphologies for two temperatures
$T=490$ K, $T=550$ K and for the rate of deposition $F_1=0.011$ MLs$^{-1}$.
Comparison of corresponding figures shows
that an excess of 1-HUC islands is well reproduced for both temperatures.

We carefully fitted the experimental data presented in Fig.
\ref{fastdep}. To simplify the fitting, we assumed
$n^*=5$ and $E_a = 0.05$ eV as obtained in
the previous work \cite{Pepa}.
We varied three parameters $n_S$, $n_B$, and $\Delta
E_S$ and at the same time we
were changing values
$E_F$ and $E_U$ within
error bars of the previous work to
find the best fit.

We found that the fitting of experimental data using the final model give values $\nu_0=$($5 \times 10^{9 \pm 1}$) s$^{-1}$ and
$E_F \approx E_U = (0.68 \pm 0.10)$ eV (at the fixed value
$\nu_0=5 \times 10^{9}$ s$^{-1}$ the values $E_F, E_U$ can be
determined with an accuracy of 0.02 eV) and the difference $E_F -
E_U = (0.03~\pm~0.01)$ eV. The values are in a good agreement with
our previous results \cite{Pepa}. In addition the model
provided energy related to effective interaction with
saturated HUC, $\Delta E_S = (0.10 \pm 0.05)$ eV, and numbers of
Ag atoms $n_S = 21 \pm 6$, and $n_B = 5 \pm 2$. Maximum number of
Ag atoms accommodated in a HUC with all three boundary areas
filled is $36 \pm 8$. It is larger than a rough estimate reported
in Ref. \onlinecite{nase} (compare also with a value of $45 \pm 5$ resulted
from the "standard" growth model). The value of $n_S$ is consistent
with a number of potential minima in a HUC -- 18 -- proposed in
Ref. \onlinecite{kaxi}.

\begin{figure}
 \includegraphics[width=8cm]{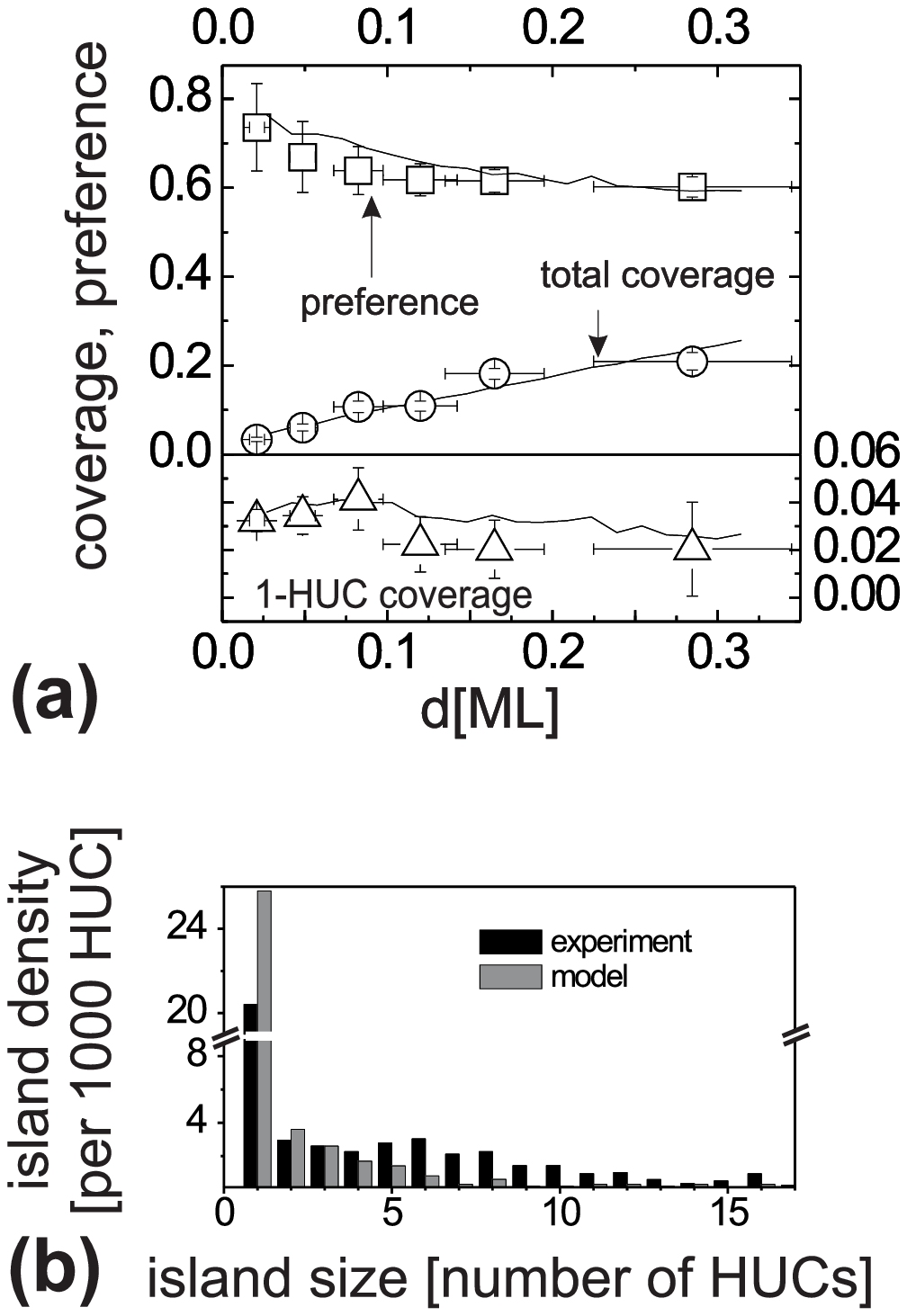}
 \caption{Comparison of experimental (symbols) and simulated (lines - best fit using the
model with centere and boundary areas) data for films
prepared at low deposition rate $F_2=0.0005$~MLs$^{-1}$ at
temperature $T=490$~K, (a) dependence of coverages and preference
on deposited amount, (b) island size distribution.
\label{slowdep}}
 \end{figure}

The  validity of the final model and the values of the parameters were
further tested for quite different growth condition - very low
deposition rate $F_2=0.0005$ MLs$^{-1}$. STM image in left part of
Fig. \ref{fdmorph}c shows that  there is
excess of 1-HUCs even in this regime.
Morphology obtained by simulation under the same
growth conditions in right part of Fig. \ref{fdmorph}c have similar
features. A series of samples provided  dependences (see Fig. \ref{slowdep}a) similar to those shown in Fig.  \ref{fastdep}. In
the same Figure, we show results obtained by calculation using the
final model with the parameters given above. There was no
additional fitting. We can see that the agreement is quite good.
The model explains well excess of 1-HUC islands  in  the size
distribution also for this much lower deposition rate
- see  a reasonable agreement between
experimental data and simulation (Fig. \ref{slowdep}b).

STM images of layers grown at various conditions show that a mean
size of large irregular Ag islands is limited. During further
growth the islands connect into a network -- the wetting layer.
The morphology of the irregular islands and wetting layer depend
on growth temperature \cite{nase,2D3D} and are driven by epitaxial
strain. The current model does not
take into account the strain which  plays an important role in the
formation of wetting layer. Therefore, the model cannot be used
for simulation of growth above 0.6 ML of deposited metal but it
allows to assess the structural properties of filling individual
HUCs.

\section{Conclusions}

Combined study of submonolayer growth of Ag on Si(111)-($7\times
7$) by series of STM measurements and kinetic Monte Carlo
simulations allowed to determine microscopic mechanism of growth
in the regime of island coalescence. The surface reconstruction
strongly affects nucleation and growth of islands. Islands do not
extend laterally but by fast connection of preformed blocks --
saturated HUCs - by filling dimer rows. Key
feature of growth is slow and {\it continuous} nucleation of new
small nuclei occupying the inner parts of HUC preferably
in proximity of already saturated HUCs.

Both surface morphologies and quantitative measurements can be
reproduced by KMC simulations using  a  model
extending the  coarse-grained model utilized
for simulation of initial stages of growth. 
It turned out that it was necessary to resolve the inner part of
HUC and its boundary.
Comparison with experiment confirmed the applicability of this  model for
interpretation of growth for a large range of deposition parameters
(temperature, deposition rate, time) and for deposited amount up to
$0.6$ ML. Correlated nucleation in proximity of saturated HUCs is
obtained by introduction of effective interaction {\it decreasing}
energy barrier for a hop into the saturated HUC by $\Delta E_S
=(0.10 \pm 0.05)$~eV. The model can be used for simulations up to
the coverage when epitaxial strain significantly influences growth
of larger islands.

The model allowed to assess
maximum numbers of Ag atoms per HUC in a 2D
island: for an isolated 1-HUC island $n_S = 21\pm 6$; for a larger
island (covering $\geq 2$ HUCs) the saturated number of Ag atoms
filling one of the three boundary areas is $n_B = 5 \pm 2 $ and
the maximum number of Ag atoms per HUC is $36 \pm 8$ ($n_S + 3
\times n_B$).

We expect that the mechanism of growth on reconstructed
Si(111)-($7\times 7$) for
other nonreactive metals is similar and that the model can be used as a starting point for modelling heteroepitaxial growth
in other systems and for fitting corresponding parameters.

\section{Acknowledgement}

This work was supported by the Grant Agency of the Czech Republic --
project 202/01/0928 and by the Ministry of Education, Youth and Sports of
Czech Republic -- project FRV\v{S} 2735/2003.

\bibliography{biblio}

\begin{thebibliography}{18}
\bibitem[{Vas()}]{Vasco1}
\bibinfo{note}{To explain this, an alternative mechanism of correlated
  diffusion [17] was suggested. Both mechanisms can give in principle
  satisfactory results. Only furher experimental data together with the
  theoretical effort can clarify this issue.}
\end{thebibliography}
\end{document}